\documentstyle[emuapj_acc,apjfonts,epsf]{article}
\def\simless{\mathbin{\lower 3pt\hbox
   {$\rlap{\raise 5pt\hbox{$\char'074$}}\mathchar"7218$}}} %< or of order
\def\simgreat{\mathbin{\lower 3pt\hbox
   {$\rlap{\raise 5pt
\hbox{$\char'076$}}\mathchar"7218$}}} %> or of order

\def\porb{{P_{\rm{orb}}}}

\def\erg{\rm {erg}} 

\def\kms{{\rm\,km\,s^{-1}}}

\def\pc3{{\rm pc^{-3}}} 
\def\msun{M_\odot} 
\def\df{\rm {df}}

\def\mmean{\langle m\rangle } 
\def\vmean{\langle v\rangle } 
\def\v2mean{\langle v^2\rangle } 

\def\mbhbig{M_{\rm {BH}}} 
\def\mbh2{m_{\rm {BH}}} 
\def\mubh{\mu_{\rm {BH}}} 
\def\bhtot{M_{\rm {BH,t}}} 
\def\mdeg{m^*_{\rm{c}}} 
\def\mcom{m_{\rm{c}}}
\def\mpul{m_{\rm{PSR-A}}} 
\def\apul{a_{\rm{PSR-A}}} 
\def\rpul{r_{\rm{PSR-A}}} 
 
\def\df{\rm {df}} 
\def\be{\begin{equation}} 
\def\ee{\end{equation}}

\def\baray{\begin{eqnarray}} 
\def\earay{\end{eqnarray}}

\begin{document}

\title{The case of PSR J1911$-$5958A in the outskirts of NGC 6752: \\ 
signature of a black hole binary in the cluster core ?}

\author{Monica Colpi\altaffilmark{1}, Andrea Possenti\altaffilmark{2}
and Alessia Gualandris\altaffilmark{1}}
\affil{\altaffilmark{1}Dipartimento di Fisica G. Occhialini, Universit\`a
degli Studi di Milano Bicocca, Piazza della Scienza 3, I-20126 Milano, Italy}
\affil{\altaffilmark{2}Osservatorio Astronomico di Bologna, 
Via Ranzani 1, I-40127 Bologna, Italy}

\bigskip

\begin{abstract} 
We have investigated different scenarios for the origin of the binary
millisecond pulsar PSR J1911-5958A in NGC6752, the most distant pulsar
discovered from the core of a globular cluster to date.  The
hypothesis that it results from a truly primordial binary born in the
halo calls for accretion-induced collapse and negligible recoil speed
at the moment of neutron star formation. Scattering or exchange
interactions off cluster stars are not consistent with both the
observed orbital period and its offset position. We show that a binary
system of two black holes with (unequal) masses in the range of
$3-100\msun$ can live in NGC6752 until present time and can have
propelled PSR~J1911-5958A into an eccentric peripheral orbit during
the last $\sim 1$ Gyr.
\end{abstract} 

\keywords{black hole physics --- globular clusters: general --- globular
cluster: individual (NGC 6752) --- pulsars: individual (PSR
J1911$-$5958A) --- stars: neutron}

\section{Introduction}

NGC6752 is a core-collapsed nearby globular cluster (GC) containing
five millisecond pulsars (MSPs), whose precise projected positions in
the sky have been recently obtained by D'Amico et al. (2002). Three of
them reside in the GC core.  In contrast, the other two pulsars are
located in the cluster outskirts. At present, they are the two outermost 
objects from the center of their host cluster
known in the catalogue of 41 pulsars in GCs
with accurately determined positions (the third ranked object being
PSR B2127+11C in M15).  The first, PSR J1911$-$5958A (PSR-A according
to D'Amico et al. 2002) is a binary MSP found at more than 3.3
half-mass radii away from the cluster center (corresponding to
$\rpul\simgreat8$ pc). Its companion is a dwarf (probably degenerate)
star of mass $\mdeg\sim 0.2\msun$, moving with period $\porb=0.86$
days on a circular orbit around PSR-A (with eccentricity $e<10^{-5}$,
D'Amico at al. 2001). The orbital separation is $\apul\sim 0.0223$ AU,
and the binary has a gravitational binding energy $E_{\rm PSR-A}\sim
10^{47}\erg.$ PSR-A is spinning with a period of 3.27 ms, has an
inferred magnetic field of $\sim 10^8$ Gauss and a characteristic age
$\tau_{\rm {PSR-A}}\sim 15$ Gyr; in short, it looks like a "canonical"
recycled MSP (Lorimer 2001). The second, PSR J1911-6000C, is located
at 1.4 half mass radii away from the core and is a MSP similar to
PSR-A.  However it lacks of a companion star and this hints for a
scenario in which PSR-C has been propelled at its current position by
a ionization event with a rare high speed star.  In this paper we will
focus our attention on PSR-A as we can
constrain its evolution from its binary nature.

PSR-A is 4 times more offset than the neutron star binary pulsar PSR
B2127+11C in M15 (Anderson et al. 1990), and has a much lighter
companion. For PSR B2127+11C, Phinney \& Sigurdsson (1991) explored an
ejection scenario in which the pulsar exchanged its light donor star
with a neutron star (NS) flying by.  Can we explain the {\it nature}
of PSR-A in similar terms?

In this paper we consider three possible explanations:
{\it (1)} PSR-A, remnant of a massive star, was 
the component of a truly {\it primordial binary} born in the GC halo or
expelled from the GC core due to the natal kick imparted by the
supernova explosion; 
{\it (2)} it was ejected from
the core via {\it scattering} or {\it exchange} interaction(s) with
cluster {\it stars}; or 
{\it (3)}  it was propelled into the cluster halo by a {\it
scattering} event involving a {\it black hole binary} (hereafter [BH+BH]). 
 
In this letter we will investigate all these three alternatives with
special consideration to the third, less intuitive, most intriguing
hypothesis. This is in part motivated by the recent discovery that
NGC6752 shows an unusual high mass-to-light ratio $\simgreat 10$ in
its inner region, corresponding to a mass $M_{\rm core}\ge 10^4~\msun$
in the form of low luminosity stellar remnants in its core (D'Amico et
al. 2002). In addition, X-ray observations (Pooley et al. 2001)
combined with upper limits on the gas content of the GC (D'Amico et
al. 2002) seem not to exclude that a significant fraction of $M_{\rm
core }$ is due to a central (perhaps binary) BH of a few hundreds
solar masses.
\vskip1.truecm

\section {Case 1: A primordial binary ?}
This  is the simplest hypothesis among the three. It requires orbital
stability against dynamical friction, and a low recoil speed at the
moment of NS formation.  Orbital stability depends on binary mass,
initial apocentric distance $r_{\rm apo}$ and orbital eccentricity,
$e_{\rm ec}.$ We computed the dynamical friction timescale
$\tau_{\df}$ for a 1/$e$ reduction
of $r_{\rm apo}$ in the case of a point mass $M$ moving in the cluster
potential (approximated as a King model) under the action of the
frictional force (Binney \& Tremaine 1987), considering $r_{\rm
apo}>r_{\rm PSR-A}$.  For a centrophobic halo orbit ($e_{\rm
ec}\lesssim 0.5$) we find that $\tau_{\df}$ is $\simgreat 10^{11}
({1.6\msun}/M)$ yr in NGC6752, implying stability of PSR-A and of its
progenitor halo binary.  If the binary has been propelled into a
highly eccentric orbit ($e_{\rm ec}\simgreat 0.9$) at the time of NS
formation, it would remain at the periphery for a time
$\tau_{\df}\simgreat 7\times 10^8 (1.6\msun/M)$ yr before returning to
the core.  Whenever $\tau_{\df}<\tau_{\rm PSR-A},$ it is the dynamical
friction time that sets the conditions on detectability.

How orbital stability combines with the problem of retention of PSR-A in
NGC6752? It is known from current proper motion measurements of 13
MSPs belonging to the Galaxy (Toscano et al. 1999) 
that MSPs have
relatively large mean transverse velocities of 85$\pm13\kms$ and  
a 3-D peculiar mean speed of $130\pm 30 \kms$ (Lyne et al. 1998;
see also Cordes \& Chernoff 1997). Theoretical
studies aimed at estimating the minimum recoil speeds $V_{\rm rec}$
(for those binary MSPs that invoke a common envelope phase preceding
the formation of the NS; Tauris \& Bailes 1996) predict for PSR-A a
recoil velocity $V_{\rm rec}\simgreat 60\kms,$ on the basis of its
observed orbital period. If the cluster escape velocity at the time of
the supernova explosion was similar to the current value $V_{\rm
esc}\sim 35~\kms$ for the center of NGC6752, the retention of PSR-A in
the cluster is indeed problematic. If the supernova took place in the
central regions, the binary of PSR-A (of mass $\simgreat 2.5\msun$
after the NS has formed) would have been driven back to the core by
dynamical friction in $\tau_{\rm df}\simless 1$ Gyr due to the high
orbital eccentricity.  As the supernova blew off in the first Gyr since
GC formation, the binary would not have survived in the halo until
present time.  PSR-A may, alternatively, descend from a binary born in
the cluster outskirts with the faintest kick ever observed
(considering that $V_{\rm esc}\sim 10 \kms$ in the halo). This
scenario calls for the NS of PSR-A being probably formed from the
accretion-induced collapse of a massive white dwarf (Grindlay et
al. 1988): in fact in this case collapse is not accompanied by severe
mass losses and this could minimize the center-of-mass runaway impulse
speed.  In this case, the time $\tau_{\rm df}$ is consistent with the
life-time of PSR-A and its evolution.

Given the estimated number $N_{\rm gal}\sim40,000$ of primordial MSPs
in the Galactic field (e.g. Lorimer 2001) and mass ratio $\mu\simless
5\times 10^{-4}$ between all known ($\sim 200$) GCs and the Milky Way,
it turns out that the overall number of MSPs born in the halo from a
primordial binary is nearly $\sim N_{\rm gal}\mu f_r\sim 3(f_r/0.15)$,
where $f_r $ corrects for the mass fraction outside radius $r$ in a
typical GC normalized to its value at $r$ equal to 3 half-mass radii.
This is a very low number considering that refers to all GCs known.
\vskip1.truecm

\section {Case 2: Scattering or exchange off a cluster's star ?} 
  
If PSR-A has been ejected out of the core by a cluster star 
the key issue is whether three-body exchange is favored
against inelastic scattering. To address this question, we need to
consider the binary nature of PSR-A and to distinguish cases in which
a dynamical interaction occurs before recycling (when the companion to
PSR-A has mass of $1<m_{\rm {c}}<2\msun$) or following it (when its
mass is $m^*_{\rm {c}}\sim 0.2\msun$).

We recall that just at the onset of recycling a "bifurcation" orbital
period $P_{\rm bif}$ separates the formation of a converging binary
(which evolves with decreasing orbital period $P_{\rm orb}$ until the
mass-losing star becomes degenerate and a short-period binary pulsar
is formed) from a diverging one (which evolves with increasing $P_{\rm
orb}$ forming a long period binary pulsar; Tauris \& Savonije 1999).
Given this scenario, PSR-A should result from a converging binary.
Thus, at the onset of recycling we use an upper limit on $P_{\rm
orb,in}<P_{\rm bif}\sim 2$ days and, as customarily, a mass of the
donor star $\sim 1-2~\msun$.

Scattering can occur either before or after recycling as it preserves,
to a high degree, the circularity seen in our binary pulsar. In both
cases, a recoil kinetic energy of $\sim 3\times 10^{46}\erg$ is
requested for ejecting [NS$_{\rm A}$+co] (comprising either PSR-A or
the not yet recycled NS and its companion) into its halo orbit.  As
this must be extracted from the [NS$_{\rm{A}}$+co] binding energy,
recoil at $V_{\rm {ej}}\lesssim V_{\rm esc}$ imposes an upper limit on
the orbital period $P_{\rm {orb,up}}$ which is $\lesssim 0.4$ days. We
infer this value using a collection of scattering events extracted
from Sigurdsson \& Phinney (1993): Table 1 gives the binary recoil
speed $V_{\rm {ej}}$ (third column) for $P_{\rm {orb,up}}$ (first line
of each case), for a period corresponding to the current binary
separation (second line), and for $P_{\rm bif}$ (third line).  Since
PSR-A is a converging system, its period can not have ever been
shorter than its current value of 0.86 days and this excludes the
scattering hypothesis.

Exchange alters the nature of the interacting system and makes the new
binary rather eccentric and with a longer orbital period.  Thus due to
the prohibitively long circularization time between compact stars,
exchange has to occur before recycling.  We consider the case (see
Table 1) of a main sequence binary [MS+ms] (MS heavier than ms)
interacting with the NS$_{\rm A}$ flying by (given the similar masses
of MS and NS$_{\rm A}$, the results are almost unaffected when
swapping their role).  We find that the target binary needs to be very
tightly bound to produce the new system [NS$_{\rm A}$+co]; after the
exchange [NS$_{\rm A}$+co] has an orbital period suspiciously close to
the currently observed value considering that recycling has not occurred
yet; characteristic times $\sim 1.8\times 10^{10}$ years for exchange
(see Table 1; first line of both exchange cases) are also found
uncomfortably long in this case.
\vskip1.truecm

\section {Case 3: Scattering off a [BH+BH] ?}

The dynamics of BHs in GCs involves repeated exchange interactions and
ejections (Sigurdsson \& Hernquist 1993; Portegies Zwart \& McMillan
2000).  This cosmic dance occurred early in the cluster lifetime among
the rich population (of few hundreds) of BHs, born from the most
massive stars: As they sink toward the center by dynamical friction
(on a time-scale $\lesssim 10^6$ yr), BH$-$[BH+BH] interactions become
overwhelmingly important. Binaries harden progressively in virtue of
these encounters and eventually leave the cluster due to recoil. As a
result, many clusters lose their original BHs on a time scale of
$\sim$ Gyr. However, dynamical studies (Portegies Zwart \& McMillan
2000; Miller \& Hamilton 2002a) show that some clusters could still
retain a binary.  We here consider the fascinating possibility that
NGC6752 belongs to this category and try to depict a scenario for the
interaction of PSR-A with a [BH+BH] system.

Two separate questions arise: ({\it i}$\,$) For how long can the
[BH+BH] binary that survived ejection avoid coalescence by emission of
gravitational waves (GWs)?  ({\it ii}$\,$) Can the [BH+BH] transfer
enough kinetic energy to propel PSR-A into the cluster halo and what
is the probability of this event?
\vskip0.5truecm
\subsection {GW Emission and Recoil in the [BH+BH]}

Here we allow BH masses in binaries to have a relatively wide spectrum
from 3 $\msun$ to $100\msun.$ The BHs may result from the direct
collapse of massive progenitor stars (Fryer 1999).  But the massive
$\simgreat 50\msun$-BHs could be the outcome of exchanges and mergers
of many ordinary ($\sim 10\msun$) BHs; in particular binary-binary
encounters can enhance their formation (Miller \& Hamilton 2002b).  As
noticed by Miller \& Hamilton (2002a), unlike the less massive BHs in
binaries which can be easily flung out by recoil before they can
merge, the more massive ones could have enough inertia to remain bound
to the cluster. Thus, the binary that avoided ejection has likely
unequal masses and could harbor the heaviest BH that NGC6752 ever had.

If we indicate with $\mbhbig$ and $\mbh2$ the masses of the two BHs in
the binary, respectively, and with $\bhtot$ and $\mubh$ their total
and reduced mass, our hypothetical [BH+BH] binary (with typical
eccentricity of $0.7$) will avoid coalescence for a time $\tau_{\rm
{GW}}\ge 10$ Gyr due to emission of GWs if its separation $a$ exceeds
a critical value $a_{\rm GW},$ where 
\be a_{\rm {GW}}\sim 0.4 \left
({\bhtot \over 100\msun}\right )^{1/2} \left ({\mubh \over
10\msun}{\tau_{\rm {GW}}\over 10^{10} \rm {yr}}\right )^{1/4}\!\!\!
\rm {AU}.
\label{eq:a_gw}
\ee 
Recoil, instead, imposes a different condition based
upon the request that the BH separation is sufficiently large to avoid
ejection: the [BH+BH] binary that remains bound to NGC6752 should
have $a>a_{\rm rec}$, where
\be 
a_{\rm {rec}}\sim
0.01 \left ({\mubh\over 10 \msun} \right)\!\!\!
\left ({m \over 10\msun} {100\msun\over \bhtot}
{35\kms \over V_{\rm{esc}}} \right)^2\!\!\!\rm {AU}.
\label{eq:a_rec} 
\ee 
Equation (\ref{eq:a_rec}) is derived knowing that binaries
interacting with much lighter intruders (of mass $m$) acquire a
kinetic energy (due to recoil) of the order of $\sim
0.2(m/\bhtot)^2\epsilon_B,$ where $\epsilon_B= G\mbhbig\mbh2/2a_{\rm
rec}$ is the binding energy of the [BH+BH] that typically produces a
recoil at the escape speed $V_{\rm{esc}}\sim 35\kms$ from the cluster
core (see Miller \& Hamilton 2002a; Hills 1992; Quinlan 1996).  In
Figure 1 we show these critical separations; when $a_{\rm GW}>a_{\rm
{rec}}$ coalescence is the correct criterion to use.

After the last encounter with a background BH, the evolution of
[BH+BH] continues; what comes into play is its interaction with the
bath of cluster stars, of mass $\mmean,$ line of sight velocity
$\vmean,$ and density $n_*$.  If $\Sigma_*\sim \,2\pi\, a G\bhtot
/\v2mean $ is the three-body interaction cross section, then the
(still very approximate) number of stars impinging onto the target
binary over a time $t\sim 10^{10}$ yr is ${\cal {{N}}_*}\sim n_*
\Sigma_*\vmean t$ and gives
\be 
{\cal {{N}}_*} \simless 2\cdot 10^3 \left ({n_*\over 10^6\rm {pc^{-3}}}
{\bhtot\over 100\msun}{6 \kms\over \vmean} 
{a\over a_{\rm {GW}}}{t\over 10^{10}\rm {yr}}\right ) 
\label{eq:n_dot}
\ee
for NGC6752. Each single star carries off an energy $\sim
0.2(\mmean/\bhtot)\epsilon_B$ so that the [BH+BH] hardens in the
course of time reaching $a=a_{\rm {GW}}$ from an initially much larger
separation $a_{\rm {acc}}\sim 2 G\bhtot/\v2mean$ (taken to be
comparable to the accretion radius) after experiencing a number of
scatterings ${\cal {N}}_{\rm {GW}}\sim 5[\bhtot/\mmean]\ln(a_{\rm
  {acc}}/ a_{\rm {GW}})$ that exceeds ${\cal {N}}_*$
\be {\cal
  {N}}_{\rm {GW}}\sim 5\cdot 10^{3}\left ({\bhtot\over
  100\msun}{0.4\msun\over \mmean}\right )\ln\left ( {a_{\rm
    {acc}}\over 80 \rm {AU}}{0.4\rm {AU}\over a_{\rm {GW}} }\right ).
\label{eq:n_gw}
\ee 
This secures a [BH+BH] lifetime comparable to that of NGC6752, or
slightly shorter.

%%%%%%%%%%%%%%%%%%% FIGURE 1%%%%%%%%%%%%%%%%%%%%%%%%%%%%
{
\vskip 0.5truecm 
\epsfxsize=9.truecm 
\epsfysize=9.truecm
\epsfbox{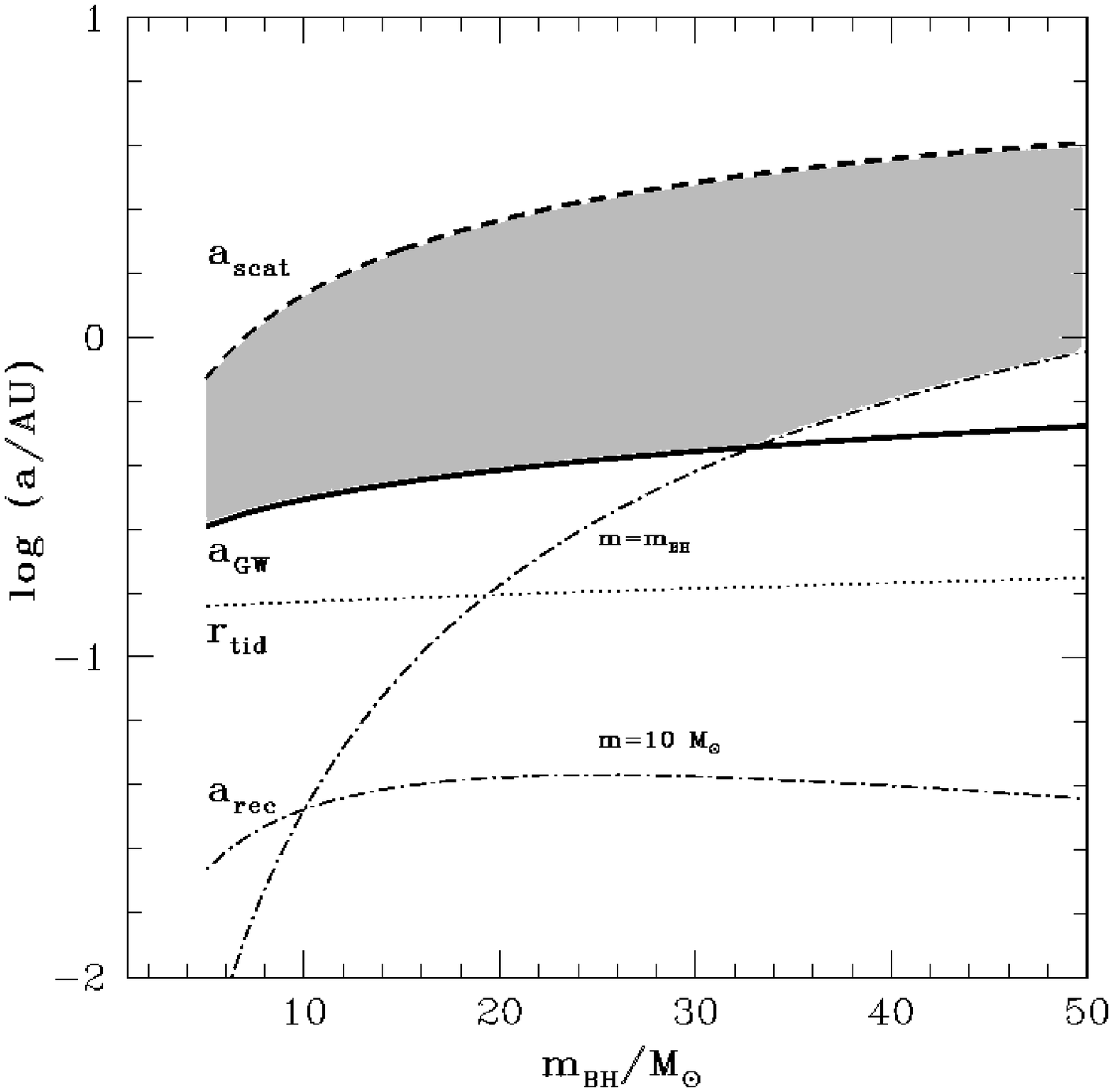} 
\figcaption[figure1.eps]{\label{fig:fig1}
\small{[BH+BH] binary separation $a$ against $\mbh2/\msun$ for
$\mbhbig=50\msun;$
$a_{\rm GW}$ (heavy solid line), and $a_{\rm{scat}}$
(heavy dashed line) are  computed from eqs.
\ref{eq:a_gw}, and \ref{eq:a_scat}. $a_{\rm {rec}},$
from  eq. \ref{eq:a_rec}, is computed for light $m=10\msun$ BH intruders and
for $m=\mbh2.$  The dotted line gives the tidal disruption radius for
[NS$_{\rm A}$+co]. The shaded area indicates the  values of
$a$ for ejection of PSR-A during the last Gyr, compatible with its detection.
}} \vskip 0.5truecm
}
%%%%%%%%%%%%%%%%%%%%%%%%%%%%%%%%%%%%%%%%%%%%%%%%%%%%%%%%%%%%%%%%%%

\subsection{PSR-A Ejection}

Let us now come to question ({\it ii$\,$}). In the ejection
hypotheses, the binary [NS$_{\rm A}$+co] must have been propelled into
an eccentric peripheral orbit at a speed $V_{\rm {ej}}\lesssim V_{\rm
{esc}}$.  The apocenter then reduces due to dynamical friction but its
$e$-folding time $\tau_{\rm {df}}$ $\simgreat 7\times
10^8(1.6\msun/M)$ yr is sufficiently long to permit the detection of
PSR-A during the pulsar life-time $\sim \tau_{\rm {PSR-A }}$.

In the {\it scattering} off [BH+BH], the light binary [NS$_{\rm
A}$+co] is viewed as a single point mass. In this case, [NS$_{\rm
A}$+co] has to avoid break up by the BH tidal field and this imposes a
constraint on the distance of closest approach, $r_{\rm min}>r_{\rm
tid}$ where 
\be r_{\rm tid} \simeq 0.2 \left({\bhtot\over 100
\msun}{1.6\msun\over \mpul+m_{\rm c}}\right )^{1/3} \!\!\!\!\!{\rm
{AU}}.
\label{eq:r_tid}
\ee 
In the inelastic scattering off the hard [BH+BH], most of the
released binding energy $\Delta\epsilon_B\sim
0.2[(\mpul+\mcom)/\bhtot]\,\,\epsilon_B$ of the binary goes into
kinetic energy of the propelled [NS$_{\rm A}$+co].  Thus for an
ejection velocity $V_{\rm {ej}}$ close to $V_{\rm {esc}},$ the [BH+BH]
should have a binary separation smaller than
\be 
a_{\rm{scat}}\sim \left ( {G\mubh\over V^2_{\rm{ej}}}\right ) 
\sim 1.3 \left({\mubh\over 10\msun} \right )
\left ({35\kms\over V_{\rm{ej}}}\right)^{2}\rm {AU}.
\label{eq:a_scat}
\ee
Since scattering leads to an effective energy exchange only when the
intruder ([NS$_{\rm A}$+co] in this case) is approaching the binary
within a distance $r$ comparable to a few orbital separation radii $a$
(beyond which energy exchange is exponentially small), conditions
(\ref{eq:r_tid}-\ref{eq:a_scat}) combined with (\ref{eq:a_gw}) or
(\ref{eq:a_rec}) give a plausible range of binary separations 
for which injection of [NS$_{\rm A}$+co] into the cluster halo is
possible and this is illustrated in Fig.~1 (shaded area).

If the flux of stars on the [BH+BH] is of thousands interactions over
the cluster lifetime (see eq.~\ref{eq:n_dot} and \ref{eq:n_gw}), about
1/30 of these encounters should have involved a NS, which gives a rate
${\cal{R}}_{ns}$ $\sim 10~{\rm Gyr}^{-1}$ (taking as face value 3,000
encounters), but it can be higher (lower) if the [BH+BH] mass is
higher (lower).  If $\sim 5\%$ of the NSs is in the state of a MSP,
then the rate for MSP ejection is ${\cal{R}}_{MSP}$ $\sim 0.5~{\rm
Gyr}^{-1},$ implying a probability of detection of about $\sim 50\%$
considering a life-time for the current orbit of $\sim 1$ Gyr.  We
note that ${\cal{R}}_{ns}$ is calculated assuming a Salpeter-like
initial mass function (giving $\sim 1\%$ of NSs) and a mass
segregation enhancement of $\sim 6$ in the density (Sigurdsson \&
Phinney 1995). If the IMF is flatter (as suggested by the observations
of D'Amico et al. 2002), this rate could be higher, enhancing the
likelihood of this scenario.  Once ejected into an halo orbit,
[NS$_{\rm A}$+co] spends only $\sim 1/50$ of its life in the dense GC
core and thus has negligible probability to undergo further
interactions (see Table~1).

\section{Discussion}
The unusual position of PSR-A in NGC6752 is puzzling.  We have shown
that PSR-A may result from a primordial binary born in the halo, but
this requires an exceptionally weak kick or accretion-induced collapse
at the moment of NS formation.  A scattering or exchange event off a
star that propelled PSR-A into the cluster halo is inconsistent with
the conservation of energy and linear momentum. Here, we propose a
dynamical interaction with a more massive system, a [BH+BH] binary in
a mass range of $\sim 3-100\msun$ and separation of $\sim$ 0.3-2 AU.
An unequal mass binary with a heavy $\simgreat 100\msun$ BH and a
lighter companion $\sim 10\msun$ is preferred, as it can form in a GC
(Miller \& Hamilton 2002a, 2002b) and can provide the right statistics for
explaining PSR-A.  Light binaries are likely to be ejected (Portegies
Zwart \& McMillan 2000), while much heavier $\simgreat 200\msun$ ones
(with equal mass) are difficult to form and their presence can be in
conflict with X-ray observations if the gas content in NGC6752 is
confirmed at the density reported in D'Amico et al. (2002); in
addition, the dynamical effects of such a heavy binary on cluster
stars could also become severe.
\footnote {A single massive $\simgreat 10^3\msun$ BH could in
principle eject PSR-A scattering off a star bound to the BH inside the
cusp (Frank \& Rees 1976) and this possibility needs to be investigated; 
note that NGC6752 has a very low line
of sight velocity dispersion and this might imply a runway growth of the
BH mass.}

Whether a GC retains its "last" BH binary(ies) or not is
unpredictable, as BH ejection depends on a number of rare random
events.  It is thus an observational challenge to unveil in the heart
of a cluster such an exotic dark binary. NGC6752 seems a primary
candidate for this search.

\acknowledgments \small {We acknowledge financial support from the
Italian Space Agency (ASI) and the Italian Minister of Research
(MIUR) under the grant MM02C71842-001.}

%%%%%%%%%%%%%%%%%%%%%%%%%%%%%%%%%%%%%%%%% TABLE 1 %%%%%%%%%%%%%%%
\begin{center}
\begin{deluxetable}{cccccc}
\small 
\tablewidth{0pt} 
\tabcolsep 0.12truecm
\tablecaption{\label{tb:tpar} {Dynamical interactions for PSR-A ejection}} 
\tablecolumns{6} \tablehead{\colhead{$P_{\rm {orb}}$}&\colhead{$a$}
&\colhead{$V_{\rm {ej}}$}&\colhead{T}&\colhead{$a'$}& \colhead{$P'_{\rm {orb}}$}\\ 
(d) & (AU) & ($\kms$) & (yr) & (AU) & (d)} 
\startdata 
\multicolumn{6}{c}{$\rm
scattering^{(b)} (1.4,1.4 MS)+(0.6 ms)\rightarrow(1.4,1.4 MS)+(0.6 ms)$}\\
\hline 
0.089 & 0.0055 & 33 & $4.6\times10^9$    & 0.0054  & 0.086\\ 
0.713 & 0.0220 & 16 & $5.8\times10^8$    & 0.0215  & 0.689\\ 
2.018 & 0.0440 & 12 & $2.1\times10^8$    & 0.0430  & 1.948\\ 
\hline 
\multicolumn{6}{c}{$\rm
scattering^{(b)} (1.4,1 MS)+(1.4 MS)\rightarrow(1.4,1 MS)+(1.4 MS)$}\\
\hline 
0.361 & 0.0135 & 34 & $2.8\times10^9$ & 0.0131  & 0.344\\
0.752 & 0.0220 & 26 & $1.4\times10^9$ & 0.0213  & 0.715\\
2.019 & 0.0425 & 19 & $5.1\times10^8$ & 0.0411  & 1.920\\
\hline 
\multicolumn{6}{c}{$\rm
scattering^{(a)} (1.4,0.5 ms)+(0.7ms)\rightarrow(1.4,0.5 ms)+(0.7 ms)$}\\
\hline 
0.067 & 0.0040 & 33 & $1.6\times10^{10}$ & 0.0039  & 0.065\\
0.868 & 0.0220 & 14 & $1.2\times10^9$    & 0.0216  & 0.843\\
2.010 & 0.0385 & 10 & $5.2\times10^8$    & 0.0378  & 1.952\\ 
\hline 
\multicolumn{6}{c}{$\rm
scattering^{(a)}
(1.4,0.14ms)+(1.4 MS)\rightarrow(1.4,0.14 ms)+(1.4 MS)$}\\ 
\hline 
0.231 & 0.0085 & 33 & $2.9\times10^{10}$ & 0.0078  & 0.205\\ 
0.962 & 0.0220 & 21 & $6.9\times10^{9}$ & 0.0203  & 0.853\\ 
2.013 & 0.0360 & 16 & $3.3\times10^9$    & 0.0332  & 1.786\\ 
\hline 
\multicolumn{6}{c}{$\rm
exchange^{(b)} (0.7 MS,0.5 ms)+(1.4)\rightarrow(0.7 MS,1.4)+(0.5 ms)$}\\
\hline 
0.382 & 0.0118 & 34 & $6.0\times10^{9}$ & 0.0330   & 1.424\\   
1.029 & 0.0229 & 24 & $2.2\times10^9$    & 0.0640   & 3.836\\ 
2.029 & 0.0360 & 19 & $1.2\times10^9$    & 0.1006   & 7.563\\ 
\hline 
\multicolumn{6}{c}{$\rm
exchange^{(b)} (1.4MS,0.6 ms)+(1.4)\rightarrow(1.4MS,1.4)+(0.6 ms)$}\\ 
\hline
0.382 & 0.0130 & 33 & $6.0\times10^{9}$ & 0.0318   & 1.2389\\
0.865 & 0.0224 & 25 & $2.7\times10^9$    & 0.0548   & 2.805\\ 
2.029 & 0.0396 & 19 & $1.2\times10^9$    & 0.0968   & 6.580\\ 
\hline \enddata 
\tablecomments{Scattering and exchanges from Sigurdsson \& Phinney
(1993); heavy (lighter) main sequence stars are denoted with MS
(ms). When indication is absent the star identify with NS$_A$ of
PSR-A.  Process labeled with {\it b} (or {\it a}) occurs {\it before }
(or {\it after}) recycling. $V_{\rm {ej}},$ $a',$ and $P'_{\rm {orb}}$
denote the ejection velocity of [NS$_{\rm A}$+co], the binary
separation and orbital period, after the encounter.  $T$ is the
interaction time scale in NGC6752 for $n=10^6{\rm {pc}}^{-3}$ (D'Amico
et al.  2002); for exchanges with incoming NSs the value of $T$ given
in the table needs to be multiplied by a factor $\sim 30,$ equal to
the expected ratio between the stellar to NS density.}
\label{tab:tab1}
\end{deluxetable}
%%%%%%%%%%%%%%%%%%%%%%%%%%%%%%%%%%%%%%%%%%%%%%%%%%%%%%%%%%%%%%%%%%%%%%%%
\end{center}
\end{document}